# A Comparative Analysis for Determining the Optimal Path using PSO and GA

Kavitha Sooda
Advanced Networking Research Group (RIIC),
DSI and Asst. Professor, Dept. of CSE
Nitte Meenakshi Institute of Technology
Bangalore, India – 64

T. R. Gopalakrishnan Nair
Advanced Networking Research Group
VP, Research, Research and Industry Incubation
Centre (RIIC), Dayananda Sagar Institutions
Bangalore, India - 78

## ABSTRACT
Significant research has been carried out recently to find the optimal path in network routing. Among them, the evolutionary algorithm approach is an area where work is carried out extensively. We in this paper have used particle swarm optimization (PSO) and genetic algorithm (GA) for finding the optimal path and the concept of region based network is introduced along with the use of indirect encoding. We demonstrate the advantage of fitness value and hop count in both PSO and GA. A comparative study of PSO and genetic algorithm (GA) is carried out, and it was found that PSO converged to arrive at the optimal path much faster than GA.

## 1. INTRODUCTION
A path, in a network, is a link between the interconnected nodes. Routing is a process of forwarding the data from a known source to the destination. In this process, the data may travel through several intermediate paths, and there exist a need to select the best possible optimal nodes to forward the data. This optimal selection of nodes will enable to achieve a high performance in the network. There are many existing work done in this area of route discovery which are discussed in the literature [1]-[2]. There may be a number of paths existing between any two nodes (source and destination), and the most efficient path in terms of the associated cost, is termed as the *shortest path* (SP). The *Shortest path problem* uses an appropriate algorithm, to obtain the shortest path in a given network. SP problems have a wide range of applications such as in transportation, communications, speech recognition and various other fields.

A number of algorithms such as Bellman-Ford [3], Dijkstra's algorithm [4], are present for the SP problem, but each has a restriction like consideration of non-negative weighted edges, determination of only one possible path between two nodes and high computational complexity for rapidly changing network. These shortcomings hence have given rise to several other SP algorithms. Artificial neural networks have also been considered but they too accompany several restrictions.

Genetic algorithm (GA) is an evolutionary computation technique which was studied and found to be successful than the ANN approach. Other evolutionary technique which can be applied for SP problem is particle swarm optimization technique [5]. Input to the PSO algorithm is given in the form of particles, hence in the case of networks, the multiple paths (between two nodes) obtained is encoded as particles. Previously for encoding paths as particles two encoding techniques were widely used, i.e. Direct Encoding and Indirect Encoding.

Here in this paper, we use a modified version of Indirect Encoding Technique and a region based network is evolved where the multiple paths between the selected two nodes which are encoded as particles. Here we have solved the shortest path problem in a region based network using PSO and compared with GA using the same fitness function. It was observed that the fitness value of the particle in PSO was better than the fitness value of the chromosome in GA. Also the hop counts obtained by PSO were lesser or equal to GA.

The next section discusses the related work, followed by the description of the region based network and Path Encoding scheme in section III. Section IV has the simulation results. The conclusion and future work are discussed in Section V.

## 2. RELATED WORK
In aiding the path selection in networks, routing algorithms play an important role. A good routing algorithm should be able to find an optimal path and it must be simple. It also must have low overhead, and be robust and stable, converging rapidly, and must remain flexible. There exists a lot of routing algorithm which have been developed for specific kind of network as well as for general routing purpose. But none of these addressed the scope involved for intelligent routing. With the ever increasing internet usage the statistical routing approach will not serve the purpose of routing faster. Lot of research work has been carried out for routing in intelligent networks. The aspect of such intelligence has been dealt in cognitive network [6] and autonomic networking [7]. Intelligent network was well formulated based on heuristic algorithm, bio-inspired computing, evolutionary algorithm and human immune system. Applying intelligent methods [8] in networks and gaining higher efficiency in operation have a greater significance in the area of optimal communication, realizing best Quality of Service and security, development of application software, and the management of different layers of protocol. The existing algorithms are either table driven or demand-driven protocols. These algorithms have been used for different applications depending on their specification. Still dynamics of nodes, hidden terminals, power aware routing; location-aid routing remains a challenge other than the QoS and multicasting. With all this limitations the protocols have been supporting the existing users of the internet. But the current trend in the usage of the internet shows that these protocols will not be self-sustainable for meeting the requirement of fast end-to-end delivery. Several algorithms have been implemented and analyzed for SP problems. In the Dijkstra's algorithm is presented for solving the shortest path problem which works for networks with non-negative weights only. For networks involving negative path network, the Bellman-Ford algorithm solves the problem, but fails in the presence of a negative weights in the networks. Similarly, [9] and [10] solves the shortest path problem for all pairs of shortest path. Although these algorithms are simple and easy to use, they have several drawbacks which make them unsuitable for practical applications.

The PSO is a method of optimizing the candidate results by iteratively trying to improvise the intermediate solution and hence obtaining the final result. Simulation studies show that the optimization method contains very few constraints. It is robust and time efficient. The cost involved in the implementation of such an algorithm is less and hence can be easily used in any real-time environment. In [11] and [12] the PSO and genetic algorithm (GA) have been compared and studied. In past several years, PSO has been successfully applied in many research and application areas and it has been proved that PSO gets better results in a faster, cheaper way compared to other methods.

## 2.1 Particle Swarm Optimization (PSO)

PSO is an optimization technique developed by Kennedy and Eberhart [13], and is inspired by the social behaviour of bird flock. As a group of birds travel certain distance, their velocity changes and accordingly the position of each of the birds also get changed, and the flock updates its position according to the position and velocity of the leading bird (or the bird which is closer to the food). Hence every bird in the flock is moving closer to the bird which is nearer to the food.

This concept of PSO has been extended to antennas, biomedical applications, communication networks, clustering and classification, combinatorial optimization, and many more areas. PSO gives us the optimal solution which always may not be the best solution, as the solution depends on the number of iterations.

PSO is used to maximize or minimize an objective function by updating the velocity (i.e. cost) in every iteration. Particles (or sequence of nodes or path) obtained by indirect encoding scheme, serve as input to the PSO algorithm. Here, our objective is to find the particle which has the maximum cost (bandwidth), associated with the links of the particle (or path).

*Map all the multiple paths from source to destination as particles.*
*Calculate pBest[] and position of corresponding particle*
*.for(i=0;i<no_of_iteration;i++)*
*for(x=0;x<particle_count;x++)*
*Change cost of each link along the path of each particle randomly.*
*Calculate fitness and find pBest[]of ith iteration.*
*Find best pBest[] i.e. gBest & corresponding particle, which gives the shortest path in terms of the cost associated with the links on the path(or particle).*

**Algorithm 1: PSO algorithm**

PSO optimizes the given particles (multiple paths of the networks) for a specified number of iterations, by considering certain parameters such as the velocity and the best position of the particle. In [5] the formulae for velocity and position have been specified. The method used is illustrated in Algorithm 1. The particle's best position is referred to as *pBest* and the best position in comparison to all other particles is referred to as *gBest*, i.e. the global best. In every iteration, the *pBest* is calculated, using which we obtain the *gBest* value, which is the shortest path of the network. The *pBest* value is calculated depending on the value of fitness of each particle in every iteration. The fitness value depends upon the bandwidth associated with the particles. The fitness equation is given below,

$$Fitness = \frac{B_{Initial\_Link}}{\sum_{i=0}^{l_{particle}} B_i} \quad (1)$$

The fitness value obtained for the particle indicates where the path can be considered for data to be sent on it. More the fitness value less the chances of corruption of data while transmission. Fitness values is also been used in GA to compare the efficiency of the approach.

## 2.2 Genetic Algorithm

In the recent past, genetic algorithm has found its application in route selection algorithms. Genetic algorithm comes under the category of evolutionary algorithm. It consists of population with chromosomes that represent the possible solutions for the given problem space. With the developments which take place in the population based on the fitness function evaluation, chromosomes are given chance to reproduce and be closer to the solution. The reproduction techniques involved are the crossover and mutation. This process is repeated for a desired number of times (iterations) to obtain the optimal solution. The importance of GA lies in the parallel working of different population. This helps to explore the complete problem space in all direction. GA is well suited where the problem space is huge and time taken to search is exhaustive. It does not require any previous knowledge to obtain the solution. The importance of GA lies in the parallel working of different populations. This helps one to explore the complete problem space in all directions. GA is well suited in cases where the problem space is huge and time taken to search is exhaustive as discussed in [14]. It does not require any previous knowledge to obtain the solution.

In both PSO and GA the fitness value of particle and chromosome has been calculated using same factors (bandwidth). Using crossovers and mutations different possible chromosomes are generated. By changing the priorities of nodes in each iteration particles position and velocity have been modified. Thus by the end of each iteration the *pBest* of each particle is been updated and at the final iteration the best particle (*gBest*) gives the shortest path from source to destination.

Here in this paper the crossover operation is performed on equal length chromosomes. We have implemented and tested for 1-point crossover technique and 2-point crossover technique. Similarly swap mutation and adjacent swap mutation is considered for simulation.

The pseudo code for GA is as follows:

k= 0;
Compute initial population $P_0$;
WHILE k <=$k_{max}$ DO
BEGIN
    Select chromosomes for reproduction;
    Create offspring's by crossover technique;
    Mutate few chromosomes for a chosen probability;
    Compute new generation;
END

The following scenario illustrates the operation of crossover and mutation of GA.

Crossover example:
   a. 1-point example:
      Parent1: **1, 2, 3, 4, 5, 6, 7, 8**
      Parent2: **1, 1, 3, 3, 4, 5, 7, 8**
      Random choice: $k = 5$
      Child: **1, 2, 3, 4, 4, 6, 7, 8**
      Child: **1, 1, 3, 3, 5, 5, 7, 8**
   b. 2-point example:
      Parent1: **1, 2, 3, 4, 5, 6, 7, 8**
      Parent2: **1, 1, 3, 3, 4, 5, 7, 8**
      Random choices: $j = 4$, $k = 6$
      Child: **1, 2, 3, 3, 5, 5, 7, 8**
      Child: **1, 1, 3, 4, 5, 6, 7, 8**

Mutation example:
   a. Swap mutation example:
      Parent: **1, 2, 3, 4, 5, 6, 7, 8**
      Random choices: $i = 3$; $j = 6$
      Child: **1, 2, 6, 4, 5, 3, 7, 8**
   b. Adjacent-swap mutation example
      Parent: **1, 2, 3, 4, 5, 6, 7, 8**
      Random choice: $j = 6$
      Child: **1, 2, 3, 4, 5, 7, 6, 8**

## 3. REGION BASED NETWORK AND PATH ENCODING SCHEME

### 3.1 Regions

A region based network is one where the entire network is spread over a number of regions, consisting of nodes as shown in Figure 1. Each region can be considered as a sub network within a city or a country. The nodes present within each region may be interconnected with each other (local nodes) or with the nodes of a different region.

In the implementation of region based approach for determining the optimal path using PSO, these regions are generated randomly based on the total number of nodes present in the network. If the number of nodes present is *pn*, then the number of regions along with the number of nodes in each region can be determined as follows,

$$pnr = \frac{pn}{a} \qquad (2)$$

where, $2^a < pn < 2^b$

Here *a* is the number of regions in the network and *pnr* is the number of nodes in a region. If the *pnr* value is a fraction, then the extra node is placed in the last region. For example, if *pn* = 21, here 24<21<25, we have 4 regions (*a* = 4).

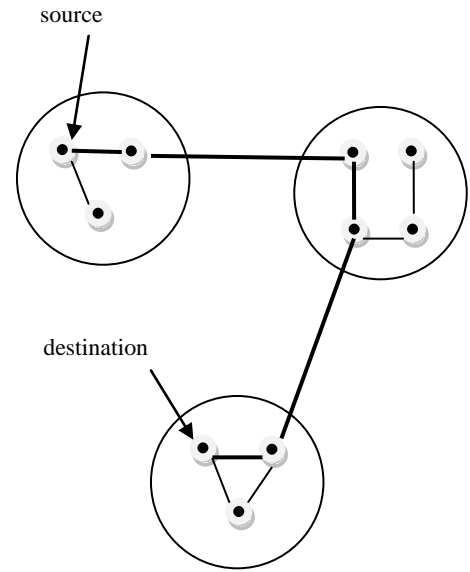

**Fig 1: Region Based network**

Now using Equation (2), 21/4=5.25 (*pnr* = 5.25). Since pnr is a fraction, each region will contain 5 nodes and the extra node (i.e. the 21st node) will be placed in the last region i.e., the fourth region in this case.

### 3.2 Path Encoding

The nodes of the network are assigned with different priorities, which are used for encoding of the path. Path construction begins at the source by considering the priorities of its neighbors. The neighbor with the highest priority is selected and added to the partial path under construction. Similarly the priorities of the neighbors of the selected node are again checked for highest priority value. This process is continued till the destination node is added in the partial path. Highest priority indicates that the node is active and the data can be sent through it without getting corrupted.

Path construction often leads to formation of loops; hence to avoid this, the selected nodes are assigned a very large negative value for their priority. In order to avoid backtracking, a *heuristic operator M* is used. The value of *M* is usually considered to be a constant value, 4 [5]. But since we are using a region based network, the *M* value has been modified and assigned the *pnr* value, i.e., *M* is now the number of nodes present in each region. It works as shown as shown below:

*if(source<destination)*
    *(ids - k)>-M*
*else*
    *(ids - k)<M*

**Algorithm 2: Heuristic operator to avoid backtracking**

In Algorithm 2, ids are the next node to be chosen in the partial path and *k* is the terminal node in the partial path.

The flow chart for the path construction is as follows:

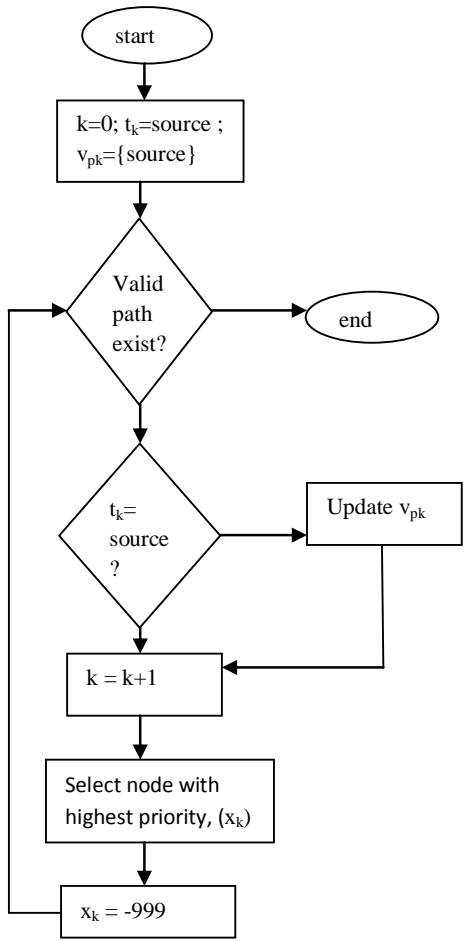

**Fig 2: Path construction**

Herein Figure 2, $t_k$ is the current or terminal node in the partial path, $v_{pk}$, under construction which contains k+1 node. Here, $x_k$ is the priority vector which contains the priority value referred to by x. Assigning -999 to the nodes selected in the partial path avoids selection of the same node again for a given path $v_{pk}$.

## 4. SIMULATION RESULTS

The topology was set up using region-based design approach. A random topology was set-up and tested for different topological structures. The regions were considered for connectivity based on the initial setup. Initially the random topology is setup. Upon this topology the path construction algorithm is applied to obtain different paths from the given source to destination. These paths which are obtained determines the input for PSO and GA. the next step is to obtain the optimal paths based on bandwidth assigned to the links using the two approaches. The results were compared for different number of iterations ranging from 5 to 20. The initial value of the iteration, i.e., from 1-5 is ignored as the values obtained did not give much clue.

Figure 3, shows the fitness value obtained by the two approaches. It was observed that the path length was with good bandwidth in PSO thus having higher value of fitness value as compared to GA.

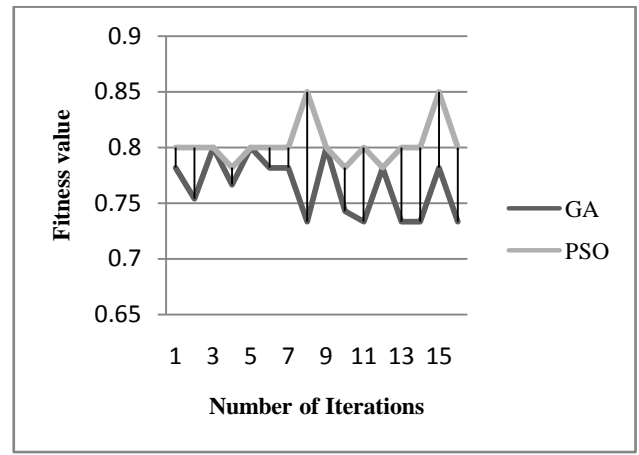

**Fig 3: Fitness value obtained by PSO and GA**

Table 1, shows the route length of paths obtained by PSO and GA. The results show that much better path was determined by PSO than GA. Here the path determination is based on the random generation of the bandwidth considered. The result in Table 1 displays for one of the randomly generated topology.

**Table 1. Comparison of PSO and GA**

|            | Hop Count ||
|------------|-----|-----|
| Iterations | PSO | GA  |
| 5          | 4   | 3   |
| 6          | 5   | 5   |
| 7          | 3   | 4   |
| 8          | 4   | 8   |
| 9          | 8   | 10  |
| 10         | 7   | 9   |
| 11         | 5   | 7   |
| 12         | 7   | 9   |
| 13         | 6   | 7   |
| 14         | 8   | 11  |
| 15         | 3   | 4   |
| 16         | 7   | 8   |
| 17         | 10  | 11  |
| 18         | 11  | 13  |
| 19         | 5   | 6   |
| 20         | 4   | 4   |

The time required for total execution of PSO and GA for various iterations under generalised normalised environment like same processor and same memory were compared and is illustrated in Figure 4. It was found that PSO converged faster than GA.

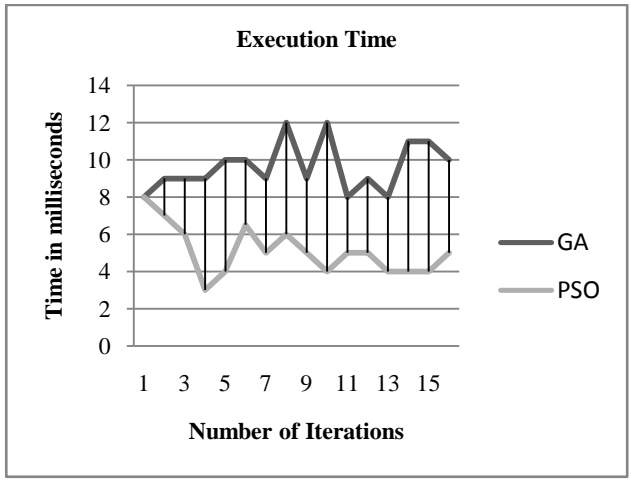

**Fig 4: Comparison for execution of PSO and GA**

## 5. CONCLUSIONS

The results obtained show a significant improvement in the convergence of optimal path by using PSO approach than GA. The region based approach helps in understanding the network design faster. The comparative results of the two algorithm show that PSO outperformed GA in obtaining better paths. As the iteration improves the particles reaches much better fitness value than the fitness of chromosome of GA. It was also observed that time taken to find the shortest path was less in case of PSO when compared to GA.

Further the algorithm maybe improved by providing knowledge about the environment of the nodes, which need to be considered to assess only optimal nodes.

## 6. ACKNOWLEDGMENTS

Our thanks to Ms. Deepthi Shetty, Ms. Prapthi Hegde and Ms. Anusha Hegde, for contributing their valuable ideas towards the work.